\documentclass[10pt,letterpaper]{article}
\usepackage{opex3}
\usepackage{graphicx}
\usepackage{amsmath}
\usepackage{bm}
\usepackage{color}

\newcommand{\un}[1]{\ensuremath{\,\mathrm{#1}}}
\newcommand{\micron}[0]{\un{\mu m}}

\newcommand{\vtr}[3]{\ensuremath{
	\left(\begin{array}{c} 
		{#1} \\ {#2} \\ {#3} 
	\end{array}\right)
}}

\newcommand{\dif}[1]{\ensuremath{\operatorname{d}\!{#1}}}

\begin{document}

\title{X-ray Fourier ptychographic microscopy}

\author{
	H. Simons$^{1,2}$,
	H. F. Poulsen$^1$,
	J. P. Guigay$^2$,
	and C. Detlefs$^{2,*}$}

\address{
	${}^{1}$Physics Department, Technical University of Denmark, 2800 Kgs. Lyngby, Denmark
  	\\ 
	${}^{2}$European Synchrotron Radiation Facility, B.P. 220, F-38043 Grenoble Cedex, France
}

\email{${}^{*}$detlefs@esrf.fr}


\begin{abstract} 
 Following the recent developement of Fourier ptychographic microscopy
 (FPM) in the visible range by Zheng et al. (2013), we propose an
 adaptation for hard x-rays. FPM employs ptychographic reconstruction
 to merge a series of low-resolution, wide field of view images into a
 high-resolution image. In the x-ray range this opens the possibility
 to overcome the limited numerical aperture of existing x-ray
 lenses. Furthermore, digital wave front correction (DWC) may be used
 to charaterize and correct lens imperfections. Given the diffraction
 limit achievable with x-ray lenses (below $100\un{nm}$), x-ray
 Fourier ptychographic microscopy (XFPM) should be able to reach
 resolutions in the $10\un{nm}$ range.
 \end{abstract}



\bibliographystyle{osajnl}
\bibliography{FourierPtychography}

\section{Introduction}

Recently Zheng et al.~\cite{Zheng2013} demonstrated Fourier
ptychographic microscopy (FPM) in the visible wavelength regime. The
technique iteratively stiches together a number of variably
illuminated, low-resolution intensity images in Fourier space to
produce a wide-field, high-resolution complex image of a
two-dimensional sample.  By varying the angle of the incident light a
wide range of scattering angles is covered -- thus improving the
space-bandwidth product (SBP) \cite{Lohmann1996} -- without moving the
sample and imaging system.

The image recovery procedure of FPM follows a stragety similar to
ptychography (that is, scanning diffraction microscopy, a technique
that is now routinely employed in the soft and hard x-ray range)
\cite{Rodenburg1992,Faulkner2004,Rodenburg2007,Thibault2008,Dierolf2010,Maiden2010,Humphry2012}:
iteratively solving for a sample estimate that is consistent with many
intensity measurements. Unlike ptycho\-graphy, however, FPM's object
support constraints are imposed in the Fourier domain, offering
several unique advantages and opportunities \cite{Zheng2013}.  

Zheng et al.~employed a conventional optical microscope with small
magnification ($\times 2$ objective), limited numerical aperture (NA)
0.08, and large field of view (FOV) $\approx 120\un{mm^2}$. Their
reconstructed FPM image had a maximum synthetic NA of 0.5
\cite{Zheng2013} set by the maximum angle between the optical axis of
the imaging lens and the illuminating beam. The resulting
reconstructed image had a resolution comparable to a conventional
microscope with $\times 20$ magnification, but the much larger FOV and
depth of field (DOF) of the low-magnification microscope. 

In the visible range, FPM is particularly useful for increasing the
FOV and DOF, as high spatial resolution can already be achieved by
using objective lenses with very large numerical aperture. In the
x-ray range, however, the resolution is limited by the small numerical
aperture of available x-ray lenses and by lens imperfections. Both of
these limitations can be addressed by FPM: The compound image
corresponds to a larger synthetic aperture, and DWC can be used to
correct for lens imperfections in data processing. The reconstruction 
yields a complex image, i.e.~both amplitude and phase contrast are 
detected. For hard x-rays, phase contrast is usually dominant and 
(non-magnified) x-ray phase contrast imaging is a growing field 
\cite{Cloetens97}.


The adaptation of FPM to x-rays is straight-forward. Rather than
changing the angle of the incident beam, however, we propose to sample
different scattering angles (and thus reciprocal space) by moving the
detector and objective lens.  


A significant difference between the visual and x-ray regimes is the
transmission profile as function of distance from the lens center
(pupil function). Lenses for visible light have a pupil function that
is completely opaque outside of the aperture, and close to 100\%
transmission throughout the active area of the lens. For x-ray lenses
the pupil function depends on the type of lens employed: For a (hard
x-ray) zone plate (ZP) \cite{Kirz1974} with dominant phase contrast
the pupil function is similar to a visible lens, whereas for a (soft
x-ray) ZP with dominant absorption contrast opaque and transparent
zones alternate. Compound refractive lenses (CRLs)
\cite{Lengeler1999}, finally, have a Gaussian pupil function
eventually terminated by an opaque limiting aperture (physical
aperture, to be distinguished from the equivalent aperture).
These characteristic pupil functions can easily be taken into account 
in the reconstruction algorithm, as we show below.

\section{Wave propagation and image formation with plane wave illumination}

The aim of the experiment is to determine the complex filter 
function $s(x_0,y_0)$ representing the sample, 
eq.~\ref{eq.sample}. For convenience, numerical efficiency 
and stability, the reconstruction is performed on the Fourier 
transform, $S'(u_0,v_0)$, see eq.~\ref{eq.Sprime}, of a phase 
shifted sample function 
$s'(x_0,y_0) = \exp\left[\frac{i\pi}{\lambda d_1} (x_0^2+y_0^2)\right] \cdot s(x_0,y_0)$, see eq.~\ref{eq.sprime}.

The (phase shifted) wave field in the detector plane, 
$g'(x_2,y_2)$, see eq.~\ref{eq.FinalImage}), is obtained 
by shifting according to the angle of the incident wave 
front (eq.~\ref{eq.ShiftWithAngle2}), multiplication with 
the pupil function (eq.~\ref{eq.PupilFunction}), and 
inverse Fourier transformation,

\begin{equation}
	g'^{(\theta,\chi)}(x_2,y_2)
	=
	{\cal F}^{-1}_{
		u_2,v_2
		\rightarrow
		x_2,y_2
	}\left[
		p(-\lambda d_2 u_2, -\lambda d_2 v_2)
		\cdot
		S'\left(
			-\frac{d_2}{d_1}u_2 - \frac{\theta}{\lambda}, 
			-\frac{d_2}{d_1}v_2 - \frac{\chi}{\lambda}
		\right)
	\right]
\end{equation} 

In order to update the Fourier map of the sample, $S'$, with data 
from the measured intensities we take the Fourier 
transform of eq.~\ref{eq.FinalImage},
\begin{eqnarray}
	G'^{(\theta,\chi)}(u_2,v_2)
	& = &
	p(-\lambda d_2 u_2, -\lambda d_2 v_2)
	\cdot
	S'\left(
		-\frac{d_2}{d_1}u_2 - \frac{\theta}{\lambda},
		-\frac{d_2}{d_1}v_2 - \frac{\chi}{\lambda}
	\right)
	\label{eq.PtychographyFinal2}
\end{eqnarray}

Eq.~\ref{eq.PtychographyFinal2} can be used to update the Fourier 
map of the sample $S'(u_0,v_0)$. 

\begin{eqnarray}
	S'_{\mathrm{new}}(u_0,v_0)
	& = &
	p^{*}(\lambda d_1 u_0+\theta d_1, \lambda d_1 v_0 +\chi d_1)
	\cdot
	G'\left(
			-\frac{d_1}{d_2}\left(u_0+\frac{\theta}{\lambda}\right),
			-\frac{d_1}{d_2}\left(v_0+\frac{\chi}{\lambda}\right)
	\right)
\nonumber \\ & & 
	+
	\left(
		1- 
		\left|
			p\left(
				\lambda d_1 u_0 + \theta d_1,
				\lambda d_1 v_0 + \chi d_1
			\right)
		\right|^2
	\right)
	\cdot
	S'_{\mathrm{old}}(u_0,v_0).\ \ \ \ \ 
	\label{eq.UpdateSample}
\end{eqnarray}

By multiplying with $p^{*}$ we remove any possible phase 
shift introduced by the pupil function and enforce the 
``support'', i.e. remove any artifacts in $G'$ outside of 
the physical aperture of the lens (see eq.~\ref{eq.support}).
The second term in eq.~\ref{eq.UpdateSample} ensures that 
information outside the pupil function is not affected by 
the update (for areas where $p=0$) and that the Fourier 
amplitude does not decay during subsequent cycles (for 
areas where $0 < \left|p\right| < 1$, which naturally occur 
with absorbing refractive lenses). Note that substituting
eq.~\ref{eq.PtychographyFinal2} in eq.~\ref{eq.UpdateSample} 
without injection of measured information into $G'$ results 
in $S'_{\text{new}}=S'_{\text{old}}$.

In the final step of the algorithm, the complex real space map 
of the sample, $s$, is obtained by inverse Fourier transform 
of $S'$ (eq.~\ref{eq.Sprime}) and removal of the phase shift, 
see eq.~\ref{eq.sprime}.
\begin{eqnarray}
	s(x_0,y_0) 
	& = &
	\exp\left[
		-\frac{i \pi}{\lambda d_1}(x_0^2+y_0^2)
	\right]
	{\cal F}^{-1}_{u_0,v_0 \rightarrow x_0,y_0} 
	\left[
		S'(u_0,v_0)
	\right]
	\\
	& = &
	\exp\left[
		-\frac{i \pi}{\lambda d_1}(x_0^2+y_0^2)
	\right]
	\iint
	\exp\left[i 2\pi(u_0 x_0 + v_0 y_0) \right]
	S'(u_0,v_0)
	\dif{u_0} \dif{v_0}
\end{eqnarray}

\section{Ptychographic algorithm}

\begin{figure*}
\centerline{
	\includegraphics[width=0.9\textwidth]{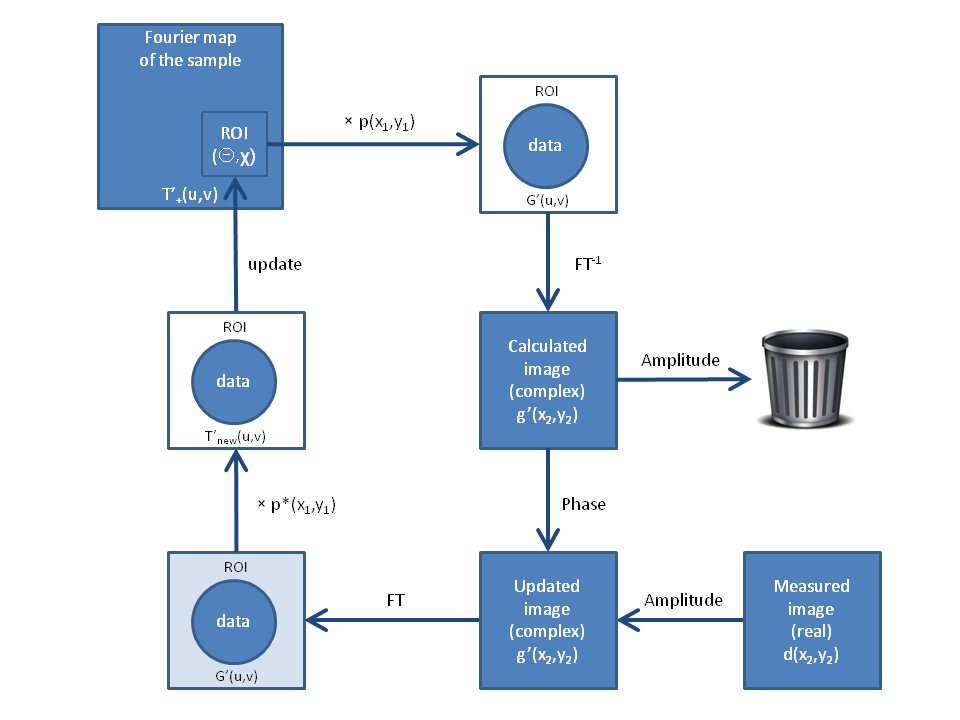}
}
\caption{\label{fig.schematic}Schematic outline of the Fourier ptychography algorithm.}
\end{figure*}

The ptychography algorithm (see Fig.~\ref{fig.schematic}) starts by 
initializing the Fourier representation of the sample, 
$S'(u_0,v_0)$ with 
more or less arbitrary values. The following steps are then repeated 
until convergence is achieved:
\begin{itemize}
\item For each pair of detection angles ($\theta, \chi$) the product 
$G'$ of the sample's Fourier map $S$ and the pupil function $p$ is 
calculated, eq.~\ref{eq.PtychographyFinal2}. As the physical 
aperture of the pupil function is significantly smaller than the
complete Fourier map of the sample these calculations are best 
carried out on a region of interest matched to the physical aperture. 
\item The complex detector field $g'(x_2,y_2)$ is obtained from $G'$ by 
inverse Fourier transform (eq.~\ref{eq.FinalImage}). 
\item The resulting wave field $g'(x_2,y_2)$
is then split into amplitude and phase. The amplitude is replaced
by the square root of the corresponding measured intensity, $d(x_2,y_2)$.
\item
The resulting ``new'' wave field $g'(x_2,y_2)$ is 
Fourier-transformed to $G'$.
\item The ``support'' is applied and possible phase shifts of the pupil 
function are removed by multiplication with $p^{*}$, see eq.~\ref{eq.UpdateSample}. 
\item The Fourier map of the sample $S'_(u_0,v_0)$ is modified using eq.~\ref{eq.UpdateSample}. 
\end{itemize}

Note that in this algorithm it makes absolutely no difference 
whether the original ($g(x_2,y_2)$) or phase-shifted 
($g'(x_2,y_2)$) image wave field is used.

\section{Digital Wavefront Correction}

Unlike classical ptychography, our treatment does not allow us 
to distinguish effects coming from the sample and from the 
incident beam. However, it should be possible to refine the 
pupil function $p(x_1,y_1)$ once the ptychographic reconstruction 
has converged, e.g. by systematically comparing the wave field 
$Q$ calculated from the measured data and the final Fourier map 
of the sample.

\section{Convergent beam illumination}

In the case of plane wave illumination discussed above, the 
reconstruction is performed on the Fourier 
transform, $S'(u_0,v_0)$, see eq.~\ref{eq.Sprime}, of a phase 
shifted sample function $s'(x_0,y_0)$.

The phase, $\Phi = \frac{pi}{\lambda d_1} (x_0^2+y_0^2)$, 
see eq.~\ref{eq.sprime}, is benign and often negligible in 
the visible range. In the hard x-ray range, however, it can 
become vicious.

A typical setup for hard x-ray microscopy using refractive lenses
could be $\lambda = 0.1\un{nm}$, $d_1 = 10\un{cm}$, 
$x_{0,\textrm{max}}, y_{0,\mathrm{max}} = 100\un{\mu m}$, 
with effective pixel size $\delta x_0 = \delta y_0 = 100 \un{nm}$.
This results in a maximum phase, in the corner of the image of
$\Phi_{\textrm{max}} = 10^3 \cdot 2\pi$, and a maximum phase 
jump between adjacent pixels of $\delta \Phi = 2\pi$ -- in the 
corner of the image, the phase of a single pixel is no longer 
well defined. The increased resolution (smaller effective pixel 
size) of the reconstructed image somewhat alleviates this problem,
but the phase drift within pixels remains problematic.

For similar experiments with Fresnel zone plates or multi-layer 
Laue lenses, the working distance $d_1$ would be even smaller, 
$d_1 ~2 \un{cm}$, such that the maximum phase and phase jump 
are proportionally higher.

It therefore appears prudent to eliminate this phase factor by 
introducing the opposite phase shift in the illuminating beam,
i.e.~by using a convergent beam that is focused onto the plane 
of the objective lens. In this case the sample function is given 
directly by $s'(x_0, y_0)$.

\section{Conclusions}

We have shown that XFPM offers many exciting possibilities for extending 
the resolution limit of full-field x-ray microscopy.

Digital wave front correction can be used to take into account the characteristic 
transmission profiles and manufacturing imperfections of x-ray lenses.

Furthermore, the phase and amplitude profiles obtained from the DWC
can be used to characterize the x-ray lens and its defects and thus
aid in optimizing the manufacturing process.

\section*{Acknowledgments} The authors thank C. Ferrero for 
stimulating discussions. We acknowledge 
the ESRF for providing financial support. 


\appendix

\section{Definitions}

All calculations are carried out in the paraxial approximation, i.e.~all 
vectors are nearly parallel to the $z$ direction. The sample, lens, and 
detector are positioned parallel to the $x$--$y$ plane at $z=0$ (sample), 
$z=d_1$ (lens) and $z=d_1+d_2$ (image/detector).

\subsection{Fourier Transform}

We denote functions in direct space by lower case letters, and
functions in Fourier space by upper case letters. 
\begin{equation}
	G(u,v) 
	= {\cal F}_{x,y\rightarrow u,v}[g(x,y)]
	=
	\iint
	\exp\left[
		-i 2\pi (u x + v y)
	\right]
	\cdot
	g(x,y)
	\dif{x}\dif{y},
	\label{eq.Fourier}
\end{equation}
and 
\begin{equation}
	g(x,y)
	= {\cal F}^{-1}_{u,v \rightarrow x,y}[G(u,v)]
	=
	\iint
	\exp\left[
		i 2\pi (u x + v y)
	\right]
	\cdot
	G(u,v)
	\dif{u}\dif{v}.
	\label{eq.invFourier}
\end{equation}
All integrals are taken from $-\infty$ to $\infty$.

\subsection{Incident wave field}

Let the incident wave field have uniform amplitude and phase, 
$t(\vec{r}) = \exp\left[i \vec{k}\cdot\vec{r}\right]$. Let the 
wave vector of the incident wave field be
\begin{equation}
	\vec{k} = \vtr{k_x}{k_y}{k_z} \approx k \vtr{\theta}{\chi}{1},
\end{equation}
where $k = 2\pi/\lambda$, $\lambda$ is the wave length, and 
$\theta,\chi \ll 1$ are the beam angles in the horizontal and vertical, 
respectively.

The sample at position $z=0$ is thus illuminated by the wave field
\begin{equation}
	t_{-}^{(\theta,\chi)}(x_0,y_0) 
	= 
	\exp\left[i k(\theta x_0 + \chi y_0)\right]
\end{equation}

\subsection{Sample}

Let the sample be represented by the complex filter function $s(x_0,y_0)$.
The wave field just downstream of the sample is thus
\begin{equation}
	t_{+}^{(\theta,\chi)}(x_0,y_0) 
	= t_{-}^{(\theta,\chi)}(x_0,y_0) \cdot s(x_0,y_0).
	\label{eq.sample}
\end{equation}

Thus with perpendicular illumination, $\theta=\chi=0$, 
the wave field just downstream of the sample is simply 
the sample filter function, 
$t_{+}^{(0,0)}(x_0,y_0)=s(x_0,y_0)$.

The Fourier transform of this field is given by
\begin{eqnarray}
\lefteqn{
	T_{+}^{(\theta,\chi)}(u_0,v_0) 
 = {\cal F}_{x_0,y_0 \rightarrow u_0, v_0}
 \left[t_+^{(\theta,\chi)}(x_0,y_0)\right]
 }
 \\
 & = &
	\iint		
	\exp\left[
		-i 2\pi (u_0 x_0 + v_0 y_0)
	\right]
	\cdot
	t_{+}^{(\theta,\chi)}(x_0, y_0)
	\dif{x_0}\dif{y_0}
\\
& = &
	\iint	
	\exp\left[
		-i 2\pi (u_0 x_0 + v_0 y_0)
	\right]
	\cdot 
	\exp\left[
		i 2\pi\left(
			\frac{\theta}{\lambda}x_0 + \frac{\chi}{\lambda}y_0
		\right)
	\right]
	\cdot
	s(x_0, y_0)
	\dif{x_0}\dif{y_0}
\\
& = &
	{\cal F}_{
		x_0,y_0 
		\rightarrow 
		u_0-\frac{\theta}{\lambda},v_0-\frac{\chi}{\lambda}
	}\left[
		s(x_0,y_0)
	\right]
\\
& = &
	S\left(
		u_0-\frac{\theta}{\lambda},
		v_0-\frac{\chi}{\lambda}
	\right),
	\label{eq.ShiftWithAngle}
\end{eqnarray}
with $S(u_0,v_0) = {\cal F}_{x_0,y_0 \rightarrow u_0,v_0}[s(x_0,y_0)]$.

Changing the angle of incidence corresponds to a shift in 
Fourier space, as noted by Zheng et al \cite{Zheng2013}.

\subsection{Lens}

Let the lens be positioned at $z=d_1$. Let the lens be described by 
the function
\begin{equation}
	l(x_1,y_1) 
	= 
	p(x_1,y_1) 
	\cdot 
	\exp\left[-ik \frac{x_1^2+y_1^2}{2f}\right],
	\label{eq.PupilFunction}
\end{equation}
where $f$ is the focal length with $1/f = 1/d_1 + 1/d_2$.

The pupil function $p(x_1,y_1)$ may be complex, e.g.~to compensate 
for a some defocussing or small lens errors. However, we require 
that its magnitude is less or equal to unity,
\begin{equation}
	\left| p(x_1,y_1)\right| \leq 1,
\end{equation}
and that it vanishes for $x_1,y_1$ outside of the physical aperture 
of the lens, e.g.~for a circular lens with physical radius $R$:
\begin{equation}
	p(x_1,y_1) = 0; \text{ for } x_1^2+y_1^2>R^2.
	\label{eq.support}
\end{equation}

This requirement provides the ``support'' for the ptychography 
algorithm by defining a limited region of interest in the Fourier 
map of the sample that affects the image for any given angle of 
illumination (and vice versa), see below.

\subsection{Image}

Let the wave field at the image plane ($z=d_1+d_2$) be $g(x_2,y_2)$. 
Evaluating the paraxial diffraction integral for the propagation 
from the sample to the lens, and then to the detector plane yields
\begin{eqnarray}
\lefteqn{
	g(x_2, y_2) =
} \nonumber \\
	& = &
	\frac{1}{i \lambda d_2}
	\iint
	\exp\left[
		\frac{i \pi}{\lambda d_2}
		\left(
			(x_2-x_1)^2 + (y_2-y_1)^2
		\right)
	\right] 
	\cdot 
	l(x_1,y_1)
\nonumber \\ & & 
	\cdot
	\left(
	\frac{1}{i \lambda d_1}
	\iint
	\exp\left[
		\frac{i \pi}{\lambda d_1}
		\left( 
			(x_1-x_0)^2 + (y_1-y_0)^2
		\right)
	\right]
	\cdot
	t_{+}(x_0,y_0)
	\dif{x_0} \dif{y_0}
	\right)
	\dif{x_1} \dif{y_1}
\\
	& = &
	\frac{-1}{\lambda^2 d_1 d_2}
	\exp\left[
		\frac{i \pi}{\lambda d_2}
		\left(
			x_2^2+y_2^2
		\right)
	\right]
	\iint
	\exp\left[
		\frac{-i 2\pi}{\lambda d_2}
		\left(
			x_2 x_1 + y_2 y_1
		\right)
	\right]
	\cdot
	p(x_1,y_1)
\nonumber \\ & &
	\cdot
	\left(
	\iint
	\exp\left[
		\frac{-i 2\pi}{\lambda d_1}
		\left( 
			x_1 x_0 + y_1 y_0
		\right)
	\right]
	\exp\left[
		\frac{i \pi}{\lambda d_1}
		\left( 
			x_0^2 + y_0^2 
		\right)
	\right]
	\cdot
	t_{+}(x_0,y_0)
	\dif{x_0} \dif{y_0}
	\right)
	\dif{x_1} \dif{y_1}
\ \ \ \ \
\\
	& = &
	-\frac{d_1}{d_2}
	\exp\left[
		\frac{i \pi}{\lambda d_2}
		\left(
			x_2^2+y_2^2
		\right)
	\right]
	\iint
	\exp\left[
		-i 2\pi
		\left(
			x_2 \frac{d_1}{d_2}u_0
			+ y2 \frac{d_1}{d_2}v_0
		\right)
	\right]
	\cdot
	p(\lambda d_1 u_0,\lambda d_1 v_0)
\nonumber \\ & &
	\cdot
	\left(
	\iint
	\exp\left[
		-i 2\pi
		\left( 
			x_0 u_0 + y_0 v_0
		\right)
	\right]
	\exp\left[
		\frac{i \pi}{\lambda d_1}
		\left( 
			x_0^2 + y_0^2 
		\right)
	\right]
	\cdot
	t_{+}(x_0,y_0)
	\dif{x_0} \dif{y_0}
	\right)
	\dif{u_0} \dif{v_0}
\ \ \ \ \ 
\\
	& = &
	-\frac{d_1}{d_2}
	\exp\left[
		\frac{i \pi}{\lambda d_2}
		\left(
			x_2^2+y_2^2
		\right)
	\right]
	\iint
	\exp\left[
		-i 2\pi
		\left(
			x_2 \frac{d_1}{d_2}u_0
			+ y2 \frac{d_1}{d_2}v_0
		\right)
	\right]
	\cdot
	p(\lambda d_1 u_0,\lambda d_1 v_0)
\nonumber \\ & &
	\cdot
	{\cal F}_{x_0,y_0 \rightarrow u_0,v_0}
	\left[
	\exp\left[
		\frac{i \pi}{\lambda d_1}
		\left( 
			x_0^2 + y_0^2 
		\right)
	\right]
	\cdot
	t_{+}(x_0,y_0)
	\right]
	\dif{u_0} \dif{v_0}
	\label{eq.intermediate}
\ \ \ \ \ 
\end{eqnarray}

For ease of notation we define a phase-shifted wave field at 
the sample position and the corresponding phase-shifted sample 
field
\begin{eqnarray}
	t'_{+}(x_0, y_0)
	& = & 
	\exp\left[
		\frac{i\pi}{\lambda d_1} 
		\left(x_0^2+y_0^2\right)
	\right]
	\cdot
	t_{+}(x_0,y_0)
\\
	s'(x_0, y_0)
	& = &
	\exp\left[
		\frac{i\pi}{\lambda d_1}(x_0^2+y_0)^2
	\right]
	\cdot
	s(x_0,y_0)
	\label{eq.sprime}
\end{eqnarray}
A change of the incident beam direction leads to a shift in 
the Fourier transform of the wave field at the sample position,
 $T'_{+}$ (eq.~\ref{eq.ShiftWithAngle}), 
\begin{eqnarray} 
\lefteqn{
	T'^{(\theta,\chi)}_{+}(u_0, v_0)
} \nonumber \\
	& = &
	{\cal F}_{
		x_0,v_0 
		\rightarrow
		u_0, v_0
	}\left[
		t'_{+}(x_0,y_0)
	\right]	
\\
	& = &
	\iint
	\exp\left[
		-i 2\pi (u_0 x_0 + v_0 y_0) 
	\right]
	\exp\left[
		\frac{i \pi}{\lambda d_1}(x_0^2+y_0^2)
	\right]
	s(x_0, y_0)
	\exp\left[ik(\theta x_0 + \chi y_0)\right]
	\dif{x_0} \dif{y_0}
	\label{eq.Tands}
	\ \ \ \ \ \ \ \ 
\\
& = &
	S'\left(
		u_0 - \frac{\theta}{\lambda},
		v_0 - \frac{\chi}{\lambda}
	\right)
	\label{eq.ShiftWithAngle2},
\end{eqnarray}
where
\begin{eqnarray}
	S'(u_0,v_0)
	& = &
	{\cal F}_{x_0,y_0 \rightarrow u_0,v_0}
	\left[
		s'(x_0,y_0)
	\right]
	\label{eq.Sprime}
	\\
	& = &
	\iint
	\exp\left[-i2\pi(u_0 x_0+v_0 y_0)\right]
	\left(
		\exp\left[
			\frac{i\pi}{\lambda d_1}(x_0^2+y_0)^2
		\right]
		s(x_0,y_0)
	\right)
	\dif{x_0} \dif{y_0}
\end{eqnarray}

Inserting these relations into eq.~\ref{eq.intermediate} 
we find
\begin{eqnarray}
	g(x_2, y_2)
	& = &
	-\frac{d_1}{d_2}
	\exp\left[
		\frac{i \pi}{\lambda d_2}
		\left(
			x_2^2+y_2^2
		\right)
	\right]
	\iint
	\exp\left[
		-i 2\pi
		\left(
			x_2 \frac{d_1}{d_2}u_0
			+ y_2 \frac{d_1}{d_2}v_0
		\right)
	\right]
\nonumber \\ & &
	\cdot
	p(\lambda d_1 u_0,\lambda d_1 v_0)
	\cdot
	T'_{+}(u_0,v_0)
	\dif{u_0} \dif{v_0}
\ \ \ \ \ 
\\
	& = &
	-\frac{d_2}{d_1}
	\exp\left[
		\frac{i \pi}{\lambda d_2}
		\left(
			x_2^2+y_2^2
		\right)
	\right]
	\iint
	\exp\left[
		i 2\pi
		\left(
			x_2 u_2
			+ y_2 v_2
		\right)
	\right]
\nonumber \\ & &
	\cdot
	p(-\lambda d_2 u_2,-\lambda d_2 v_2)
	\cdot
	T'_{+}\left(
		-\frac{d_2}{d_1} u_2,
		-\frac{d_2}{d_1} v_2
	\right)
	\dif{u_2} \dif{v_2}
\ \ \ \ \ 
\\
	& = &
	-\frac{d_2}{d_1}
	\exp\left[
		\frac{i \pi}{\lambda d_2}
		\left(
			x_2^2+y_2^2
		\right)
	\right]
\nonumber \\ & &
	\cdot
	{\cal F}^{-1}_{
		u_2,v_2
		\rightarrow
		x_2,y_2
	} \left[
		p(-\lambda d_2 u_2, -\lambda d_2 v_2)
		\cdot
		T'_{+}\left(
			-\frac{d_2}{d_1} u_2,
			-\frac{d_2}{d_1} v_2
		\right)
	\right],
	\label{eq.ImageField}
\end{eqnarray}

We have thus obtained a relation between the observed wave field, 
$g$, and the Fourier transform of the sample, $S$, via the
phase-shifted field $T'_{+}$. 

The detector is sensitive only to the intensity,
$\left|g(x_2,y_2)\right|^2$, therefore the phase 
factor $\exp\left[\frac{i \pi}{\lambda d_2}(x_2^2+y_2^2)\right]$ 
in eq.~\ref{eq.ImageField} does not influence the measurements. 
We absorb it, together with the constant amplitude factor 
$-d_2/d_1$ due to the magnification of the image into the 
effective field $g'(x_2, y_2)$.
 
\begin{eqnarray}
	g'(x_2,y_2)
	& = & 
	-\frac{d_1}{d_2}
	\exp\left[
		\frac{-i \pi}{\lambda d_2}
		\left(
			x_2^2 + y_2^2
		\right)
	\right]
	\cdot
	g(x_2,y_2)
\\
& = &
	{\cal F}^{-1}_{
		u_2,v_2
		\rightarrow
		x_2,y_2
	}\left[
		p(-\lambda d_2 u_2, -\lambda d_2 v_2)
		\cdot
		S'\left(
			-\frac{d_2}{d_1}u_2 - \frac{\theta}{\lambda}, 
			-\frac{d_2}{d_1}v_2 - \frac{\chi}{\lambda}
		\right)
	\right]
	\label{eq.FinalImage}
\end{eqnarray} 
Eq.~\ref{eq.FinalImage} is our final result for calculating the 
image from the product of the direct space pupil function, $p$ 
(with suitably scaled arguments), and the Fourier transform of 
the phase-shifted sample field, $S'$.

\section{Array sizes, units and pixels}

The units of all real space variables ($x,y$) are in meters while all 
Fourier space variables ($u,v$) are in inverse meters. Angles are 
in radians
and wave fields in both real and Fourier space are unitless. The pupil 
function is also unitless.

Array sizes are estimated as follows (for simplicity we assume 
$x$ and $y$ to be identical):

\begin{itemize}
\item Let the detector (represented by the array $d$) have 
$N_d$ pixels of size $\Delta x_2 = a$, giving a field of 
view $x_{2,\mathrm{max}} = N_d a$. 

Typical values are $N_d =1000$ and $a=1\micron$.

\item The complex detector field $g'$ has to be compared 
directly to the detected amplitude $d$. Arrays size and 
pixel size of $g'$ and $d$ should therefore be identical, 
$N_g = N_d$.

\item The FT of $g'$ has pixel size 
$\Delta{u_2} = 1/x_{2,\mathrm{max}} = 1/(N_g a)$. 
As $Q(\mu u_2, \mu v_2)$ is the {\em scaled} FT of $g'$, 
the array $Q$ has pixel size 
$\Delta u_0 = \Delta (\mu u_2) = \mu/(N_g a)$ and field 
of view $u_{0,\mathrm{max}} = \mu u_{2,\mathrm{max}} = \mu/a$. $N_Q=N_g$.

Typical values (for a typical magnification $\mu = 10$) are 
pixel size $\Delta u_0 = \Delta (\mu u_2) = 10^4\un{m^{-1}}$ 
and thus field of view $u_{0,\mathrm{max}}(\mu u)_{2,\mathrm{max}}=10^7\un{m^{-1}}$.

\item The array $p$ has pixel size $\Delta x_1 = \Delta (\mu u) d_1 \lambda = \mu d_1 \lambda/(N_g a)$. $N_p = N_Q = N_g$

A typical value is ($d_1=0.33\un{m}$) $3.3\cdot 10^{-7}\un{m}$, 
with field of view $330\micron$ (compared to typical Be lenses 
with effective apertures of $R\approx 300\micron$).

\item Angular shifts should be $\theta$ several times 
the physical aperture $R$ divided by the sample-lens 
distance $d_1$,
$\theta \approx M \cdot R/d_1 \approx M\cdot 3.3^{-4}\un{m}/(0.33\un{m})= M \cdot 1\un{mrad})$. 

The corresponding shift in Fourier 
space is $M \cdot \theta/\lambda \approx M \cdot 10^{-3}/(10^{-10}\un{m})=M \cdot 10^7\un{m^{-1}}$ 
(consistent with the value for the field of view found above).

\item Adding the shift in Fourier space to the field of 
view of $Q$ yields the full size of the Fourier map of 
the sample, $S$, i.e. $(M+1)\times$ the Field of view 
of $Q$ for a shift by $M$ times the physical aperture $R$.

As the pixel size in $S$ is the same as in $Q$, the number of 
pixels in $S$ has to be $N_s = (M+1) N_g$.

\item The corresponding resolution of the sample in real 
space is $1/(N_s \Delta(\mu u)) = a/((M+1) \mu)$.    
\end{itemize}

\end{document}